\documentclass{article}
\usepackage[a4paper,tmargin=30mm,bmargin=30mm,lmargin=15mm,rmargin=15mm,headsep=20mm]{geometry}

\usepackage{amssymb}
\usepackage{url}
\usepackage[hidelinks=true]{hyperref}
\usepackage{multirow}
\usepackage{csquotes}
\usepackage{graphicx}
\usepackage[acronym, shortcuts]{glossaries}

\makeglossaries

\newacronym{ae}{AE}{Autoencoder}
\newacronym{ai}{AI}{Artificial Intelligence}
\newacronym{ann}{ANN}{Artificial Neural Network}
\newacronym{cbir}{CBIR}{Content-Based Image Retrieval}
\newacronym{cdss}{CDSS}{Clinical Decision Support System}
\newacronym{cnn}{CNN}{Convolutional Neural Network}
\newacronym{cv}{CV}{Computer Vision}
\newacronym{dl}{DL}{Deep Learning}
\newacronym{ed}{ED}{Pleritumoural edema}
\newacronym{et}{ET}{Gd-enhancing tumor}
\newacronym{ml}{ML}{Machine Learning}
\newacronym{mr}{MR}{Magnetic Resonance}
\newacronym{mri}{MRI}{Magnetic Resonance Imaging}
\newacronym{mocae}{MOC-AE}{Multi-Output Classification Autoencoder}
\newacronym{net}{NET}{Non-enhancing tumor core}
\newacronym{nifti}{NIfTI}{Neuroimaging Informatics Technology Initiative}
\newacronym{nn}{NN}{Neural Network}
\newacronym{relu}{ReLU}{Rectified Linear Unit}
\newacronym{svm}{SVM}{Support Vector Machine}
\newacronym{t1}{T1}{Native scanner}
\newacronym{t1gd}{T1Gd}{Post-contrast T1 weighted}
\newacronym{t2}{T2}{T2 weighted}
\newacronym{flair}{T2-Flair}{T2 Fluid Attenuated Inversion Recovery}
\newacronym{vae}{VAE}{Variational Autoencoder}
\newacronym{brats}{BraTS}{Multimodal Brain Tumor Segmentation Challenge}

\begin{document}

\title{Artificial Intelligence Model for Tumoral Clinical Decision Support Systems}

\author{Guillermo Iglesias
\thanks{Departamento de Sistemas Informáticos, Escuela Técnica Superior de Ingeniería de Sistemas Informáticos, Universidad Politécnica de Madrid, Spain}
\\{guillermo.iglesias@upm.es (Corresponding author)}
\and
Edgar Talavera
\footnotemark[1]
\\{e.talavera@upm.es}
\and
Jesús Troya
\thanks{Infanta Leonor University Hospital. Madrid. Spain}
\and
Alberto Díaz-Álvarez
\footnotemark[1]
\\{alberto.diaz@upm.es}
\and
Miguel García-Remesal
\thanks{Biomedical Informatics Group, Departamento de Inteligencia Artificial, Escuela Técnica Superior de Ingenieros Informáticos, Universidad Politécnica de Madrid, Spain}
\\{mgremesal@fi.upm.es}
}

\date{}

\maketitle

\begin{abstract}
\textbf{Background and Objective:}
Comparative diagnostic in brain tumor evaluation makes possible to use the available information of a medical center to compare similar cases when a new patient is evaluated. By leveraging Artificial Intelligence models, the proposed system is able of retrieving the most similar cases of brain tumors for a given query. The primary objective is to enhance the diagnostic process by generating more accurate representations of medical images, with a particular focus on patient-specific normal features and pathologies. A key distinction from previous models lies in its ability to produce enriched image descriptors solely from binary information, eliminating the need for costly and difficult to obtain tumor segmentation.

\textbf{Methods:}
The proposed model uses Artificial Intelligence to detect patient features to recommend the most similar cases from a database. The system not only suggests similar cases but also balances the representation of healthy and abnormal features in its design. This not only encourages the generalization of its use but also aids clinicians in their decision-making processes. This generalization makes possible for future research in different medical diagnosis areas with almost not any change in the system.

\textbf{Results:}
We conducted a comparative analysis of our approach in relation to similar studies. The proposed architecture obtains a Dice coefficient of 0.474 in both tumoral and healthy regions of the patients, which outperforms previous literature. Our proposed model excels at extracting and combining anatomical and pathological features from brain \glspl{mr}, achieving state-of-the-art results while relying on less expensive label information. This substantially reduces the overall cost of the training process. Our findings highlight the significant potential for improving the efficiency and accuracy of comparative diagnostics and the treatment of tumoral pathologies.

\textbf{Conclusions:}
This paper provides substantial grounds for further exploration of the broader applicability and optimization of the proposed architecture to enhance clinical decision-making. The novel approach presented in this work marks a significant advancement in the field of medical diagnosis, particularly in the context of Artificial Intelligence-assisted image retrieval, and promises to reduce costs and improve the quality of patient care using Artificial Intelligence as a support tool instead of a black box system.
\end{abstract}


%% \linenumbers
\glsresetall
\section{Introduction}
The techniques used in medicine to treat pathologies are becoming more advanced each day and can perform less invasive and more efficient treatments~\cite{he2019practical}. However, a correct disease diagnosis is necessary before a patient can apply proper treatment, which becomes an important task in this process~\cite{yanase2019systematic}.

Although medicine has radically evolved, diagnosis is still mostly a human process. The expert must be able to thoroughly evaluate the patient's evidence and avoid making mistakes during the process~\cite{jussupow2021augmenting}. A late or incorrect diagnosis can lead to an increase in pathologies that, in cases such as cancer, can be fatal and irreversible~\cite{tan2006early, ragab2022ensemble}.

Therefore, any improvement that a physician can make during the diagnostic process may be crucial and necessary to greatly improve treatment results, because early diagnosis improves treatment results, as it improves the result of the procedure~\cite{kimber2021potential, sullivan2021earlier, haq2021deep}. Currently, computer algorithms are specially relevant, since they can be used as an additional tool for decision making, known as \gls{cdss} \cite{rana2023machine, ahmad2022comparative}. It is precisely in the current paradigm of health information science where these \glspl{cdss} can be used to process in real time large amounts of data~\cite{sutton2020overview}, using the available information to ease the physician's diagnostic process.

In this sense, there are two approaches to processing large amounts of medical images, \gls{cbir} and the so-called Concept-Based Image Indexing systems. The latter use meta-information of the images to extract their content and recommend similar samples from the database.

\gls{cbir} systems are focused on processing large amounts of data and \gls{ai} algorithms arise as the best solution to this problem. These frameworks can obtain the most similar images from a dataset given a certain query. Different solutions have been proposed in recent years, combining \gls{cbir} with \gls{dl} techniques~\cite{siradjuddin2019feature, kobayashi2021decomposing}. These systems make it possible to use large amounts of information in an ordered manner, taking advantage of the information available without using it manually \cite{karar2022intelligent}. The lastest studies in this area suggest that \gls{dl} techniques outperform previous traditional algorithms in diagnosis \cite{rana2023machine}.

However, it is not widely considered that \gls{ai} should replace health professionals in tasks they normally perform~\cite{homolak2023opportunities} but instead assist them in diagnosis and decision making, considering that the human professional will always make the final decision. Aspects such as lack of explainability \cite{ahmad2022efficient} or precision~\cite{ossa2022re}, or inability to deal with outliers suggest that it is probably too early to completely delegate critical tasks to an AI, at least in the foreseeable future~\cite{krittanawong2018rise}.

There are efforts to tackle the main problems of \gls{ai}, such as the lack of explainability using Explainable \gls{ai} with works such as \cite{du2022explainable}. In this sense the most relevant literature \cite{hauser2022explainable, albahri2023systematic, wang2023improved} does not seek to provide a standalone diagnostic of a certain case, but to provide tools to the physician to help in their diagnostic. \gls{cdss} arises as other solution to help the physician to use the latest advances in \gls{ai}.

One of the techniques in which \gls{ai} has recently achieved better results is in \gls{cv}~\cite{guo2022attention}. Here, these algorithms can use medical imaging data to extract and process its information to help professionals perform certain tasks~\cite{esteva2021deep, ward2021computer}. Thus, this relationship between \gls{cv} and medical imaging has led to the emergence of works combining the latest advances of \gls{ai} in image feature extraction and  medical data~\cite{ji2021learning, karimi2021convolution}.

\glspl{ae}~\cite{rumelhart1985learning} are machine learning models in dimensionality reduction and feature extraction processes. In this work, it is proposed to use a novel \gls{ae} variant, that factorizes brain tumoural images, to use the generated image descriptors to recommend cases with a similar pathology.

The presented model focuses on improving the retrieval accuracy of a standard \gls{ae} without using costly information as previous similar state-of-the-art approaches, such as tumor area segmentation~\cite{kobayashi2021decomposing}. Our contribution can balance the normal anatomical features of each patient, i.e. the healthy regions, with the tumor features in a single descriptor. Thus, the model can recommend interesting cases considering relevant medical information, such as the tumor area, position in the brain, composition and geometry.

Different works published in the last decade use machine learning techniques to improve the doctors' diagnoses~\cite{fatima2017survey, ralbovsky2020towards, haq2021deep, quellec2010wavelet}. Using the latest \gls{ai} research and applying it to the medical field, different works have obtained impressive results in tasks related to the medical field~\cite{gardezi2019breast}.

Regarding medical imaging algorithms, \glspl{cdss} are a prolific area where, in recent years, many articles have been published~\cite{musen2021clinical, rani2021decision, rama2022role, tuppad2022machine}. These techniques combine the potential \gls{ai} can achieve in extracting the most important features of medical images and support systems that provide the doctor with the most relevant information in each case~\cite{jiang2016scalable}.

Respect oncology, there is also different works published in the last years focusing on using the knowledge base available to perform comparative diagnostic \cite{wang2023artificial}. The main drawback of these methods is the necessity of structured information for use this knowledge base, where the image descriptors formed by \gls{cv} algorithms can produce great results.

The present work presents a novel and efficient approach to \acrfull{cbir} in medical diagnosis using \acrfull{ai} models with the ability to generate an enriched image descriptor using only binary label information, bypassing the need for costly tumor segmentation and using only binary labels, enhancing the diagnostic process by providing a more accurate representation of medical images, focusing on the normal features and pathologies in patients.

\section{Methods}

\subsection{\acrlong{brats} dataset}
\label{section:Dataset}
We used \gls{brats} 2020 dataset~\cite{menze2014brats1, bakas2017brats2, bakas2018brats3} which contains 369 labeled \gls{mr} cases of gliomas. The available data is manually segmented obtaining the tumoural region of each case in a separate file. Each case has a resolution of $240\times240\times155$ and is available in different sequences. We used this database as a baseline for the proposed Information Retrieval system, because it presents real case \glspl{mr} which has been tested before in similar systems.

The dataset is divided in two partitions: \textit{MICCAI\_BraTS\_Training} contains information on 369 different cases; this subset also contains manually segmented regions of the tumoural areas of each case. Additionally, \textit{MICCAI\_BraTS\_Validation} contains 125 non-labeled \gls{mr}.

Each \gls{mr} is available as four different sequences: \gls{t1}, \gls{t1gd}, \gls{t2} and \gls{flair}. Segmentation of each \gls{mr} is divided into three labels, \gls{et}, \gls{ed} and \gls{net}, manually segmented and approved by neuroradiologists.

Due to the fact that for the training and evaluation of the proposed algorithm, the labeled information for each  \gls{mr} is necessary, it will only be used in the \textit{MICCAI\_BraTS\_Training} partition. This dataset will be divided into a training and evaluation set.

\begin{figure}[h]
	\centering
	\includegraphics[width=\linewidth]{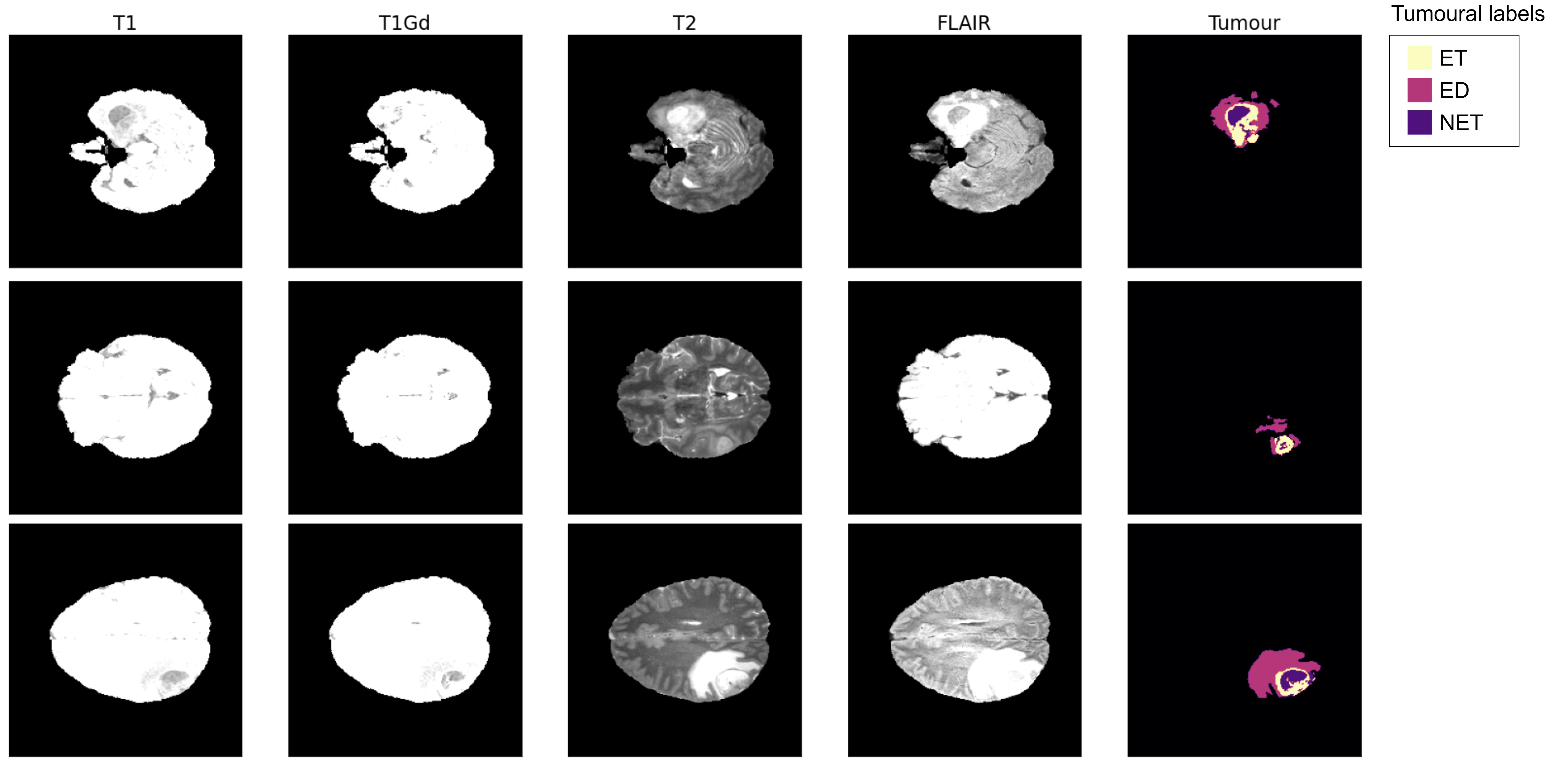}
    \caption{Sample images from \gls{brats} dataset. Each row contains a different patient and each column a different information for the patient, from \gls{mr} to tumour segmentation information.}
    \label{figure:BratsDataset}
\end{figure}

Figure \ref{figure:BratsDataset} contains samples of the information present in the dataset. The first 4 columns have a different sequence, denotated by a text in the top of the column and the 5 column has the tumor segmentation labels of the case. It must be noted that each image represents a full three dimensional \gls{mr} of a patient but only a section of the full \gls{mr} is showed.

To test the cases, we study the similarity of segmentation between the query and the retrieved images will be studied. In the tumoural characteristics of each patient, similarity measurement is performed using the tumor segmentation information present in the original dataset. However, to compare the similarity of the healthy areas of each patient, we have the same approximation of~\cite{kobayashi2021decomposing} where the brain of each patient is divided anatomically. This process must be performed to compare the results of the proposed model with those of the work of~\cite{kobayashi2021decomposing} and, to obtain the same information they used, each \gls{mr} must be preprocessed.

\subsubsection{Dataset preprocessing}
\label{section:DataPreprocessing}
Each three dimensional \gls{mr} consists of $240\times240\times155$ pixels of information, but the input of the proposed method is a two dimensional image. To obtain the images from the three dimensional data that are stored in the \gls{brats} 2020 dataset the \glspl{mr} must be sliced in layers. The data dimension of $240\times240\times155$ is sliced on the third axis to generate $240\times240$ images by taking the information about each layer separately. Furthermore, each image is normalized in the range [-1, 1], to then be properly treated with \glspl{ann}. The normalization process can be defined by the following equation:
\begin{equation}
\begin{split}
x'_i = \frac{x_i}{127.5} - 1 
\\
\forall x'_i \ \ \epsilon \ \ [-1,1]
\end{split}
\end{equation}

where $x$ denotes the gray level of each pixel of the image, where $x \epsilon [0, 256]$.

In addition, each healthy image is labeled with six normal anatomical labels: left and right cerebrum, left and right cerebellum and left and right ventricle. This division is achieved using BrainSuite 19a software~\cite{shattuck2002brainsuite}. This program can obtain a voxel segmentation of each case's cerebrum, cerebellum and ventricle, making it possible to use this information to evaluate the similarity between the query and the retrieved cases.

\begin{figure}[h]
    \centering
    \includegraphics[width=0.7\linewidth]{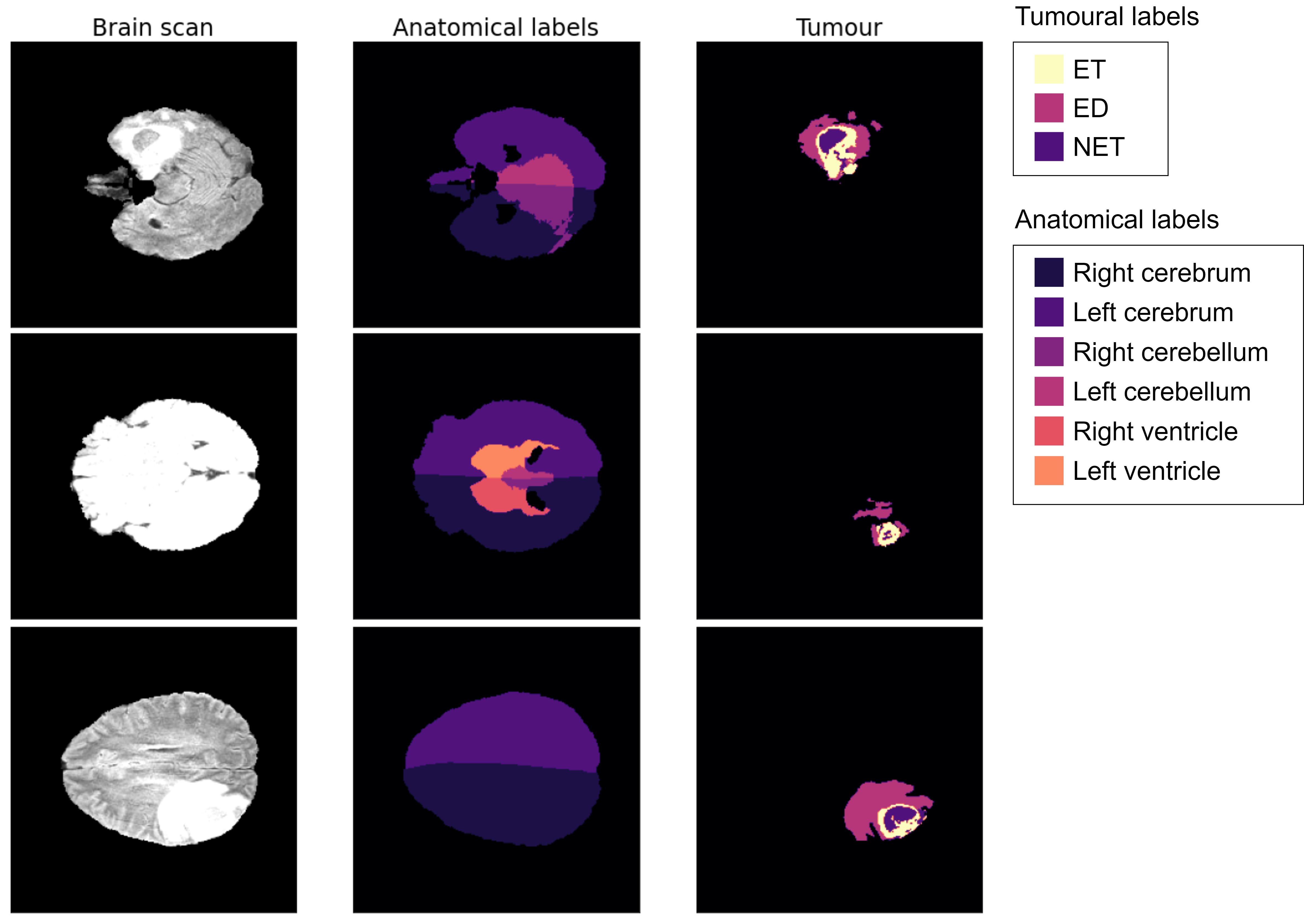}
    \caption{Sample images of the labeled dataset. Each row contains a different case and each column contains the patient \gls{mri}, anatomical label segmentation and tumoural segmentation.}
    \label{figure:BratsSegDataset}
\end{figure}

Figure \ref{figure:BratsSegDataset} has a sample of the new labels generated for each image. As seen after segmentation of the anatomical labels, each brain is divided into six different areas, as was done in \cite{kobayashi2021decomposing}. The second column of the figure has the generated anatomical labels for each section of the patient, whereas the first column is the \gls{mri} and the third one the tumour segmentation information.

\begin{figure}[h]
	\centering
	\includegraphics[width=\linewidth]{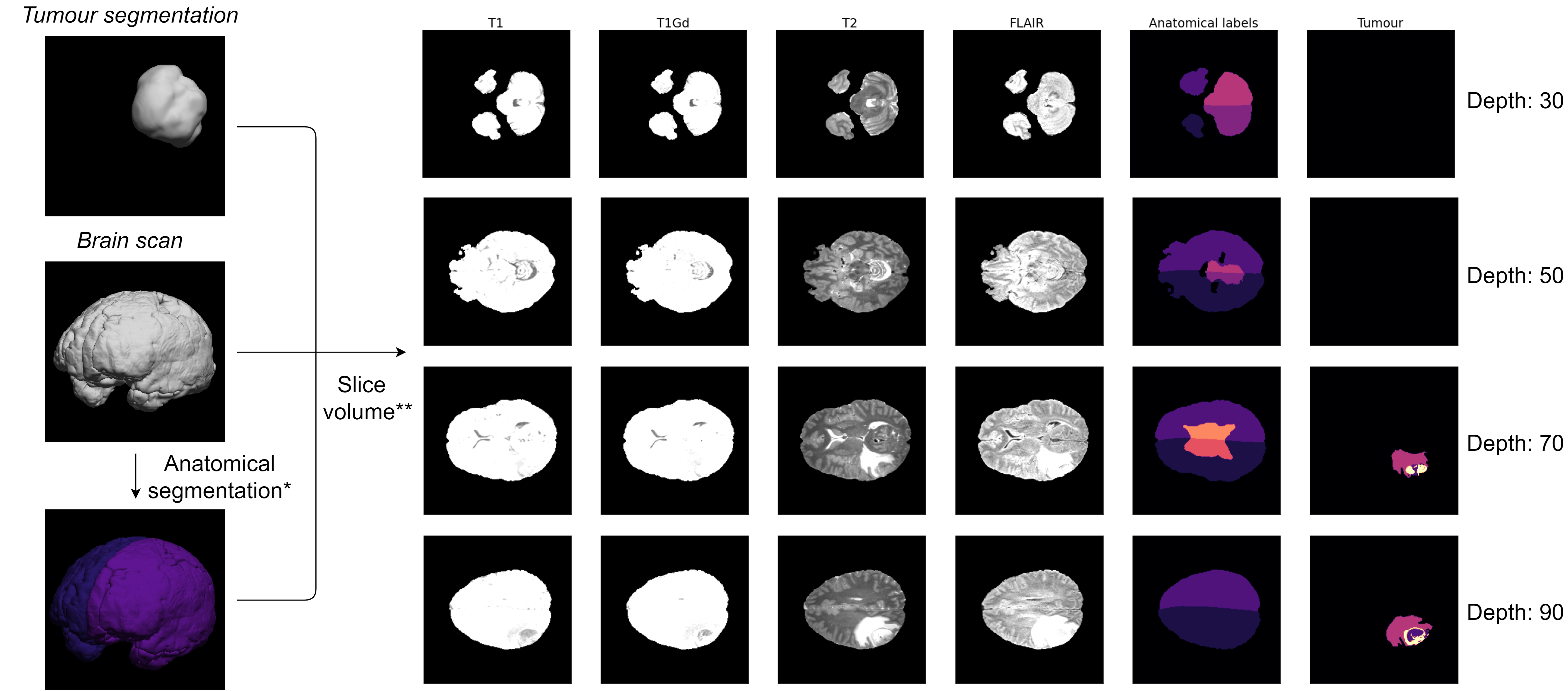}
    \caption{Data preparation scheme. (*) show that the anatomical labels were obtained using BrainSuite 19a software~\cite{shattuck2002brainsuite}. (**) shows that each slice corresponds to a certain depth in z axis.}
    \label{figure:Bratspreprocess}
\end{figure}

Figure \ref{figure:Bratspreprocess} shows a brief scheme of the preprocessing process, from the 3-dimensional \gls{nifti} files to a two dimensional images of each case, obtaining in addition a segmentation of the anatomical labels of each patient using BrainSuite 19a software \cite{shattuck2002brainsuite}. Each row represent a different slice of the same case.

\subsection{Architecture definition}
\label{section:mocae}
\begin{figure}[h]
	\centering
	\includegraphics[width=0.7\linewidth]{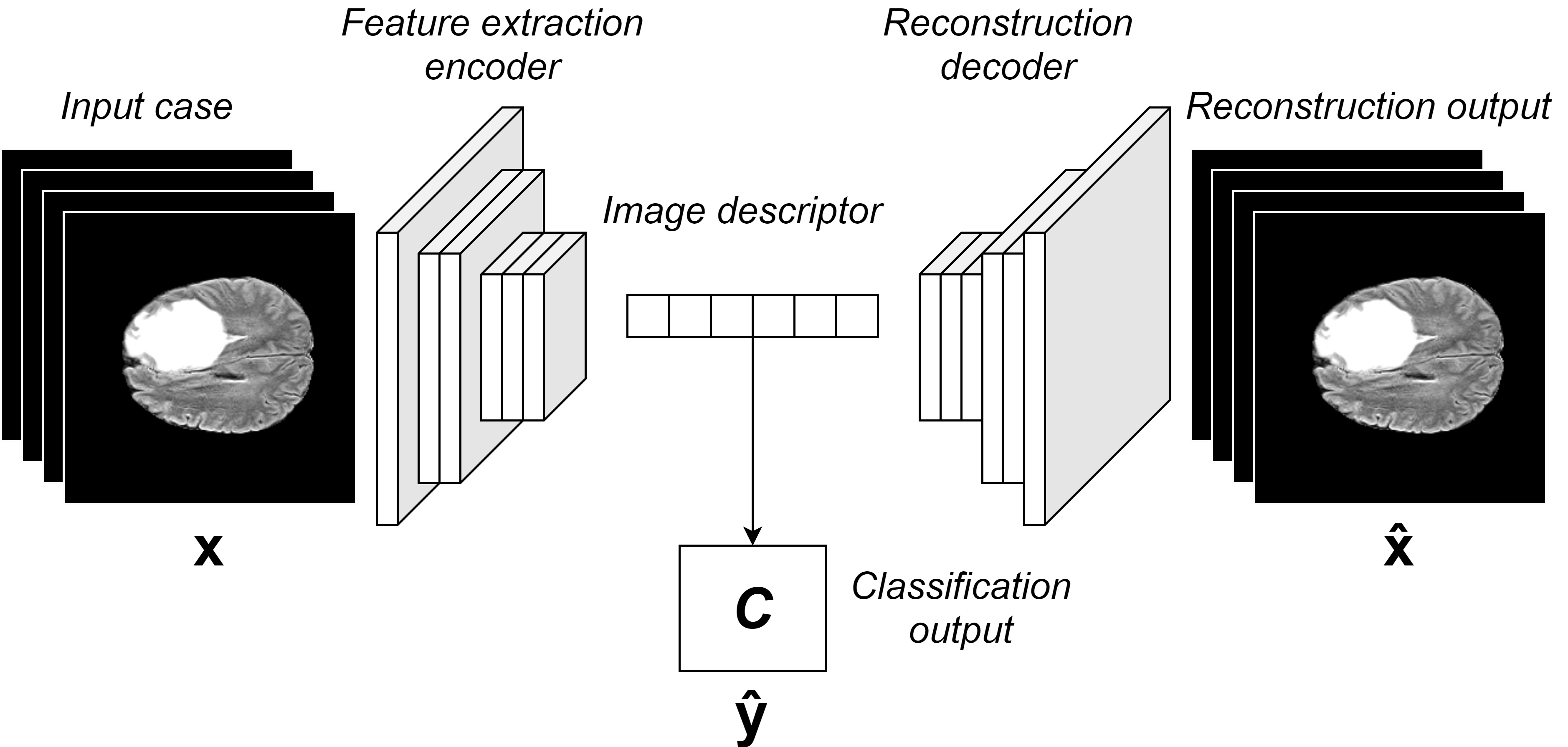}
    \caption{Neural network scheme of the proposed model.}
    \label{figure:Architecture}
\end{figure}

In Figure \ref{figure:Architecture} the schematic of the proposed method, named as \gls{mocae}, can be seen. According to the figure, the architecture presented combines two approaches: an \gls{ae} network that extracts the structural information of each image and a binary classifier that is responsible for extracting the tumor information from each case, i.e. if a tumor is present or not in the image received. This dual-objective architecture enhances the features represented in the descriptor obtained from the latent vector.

On the one hand, in \gls{mocae} an \gls{ae} is used following the same approximation as~\cite{siradjuddin2019feature} where a \gls{cbir} system is designed using the latent space of an \gls{ae} as image descriptors. This simple scheme makes it possible to extract the composition features of an image by forcing the network to reduce the dimensionality of the input images. One of the main advantages of this method is that it does not require any label to work, resulting from the self-supervised learning scheme of \glspl{ae}~\cite{krishnan2022self}. Using an \gls{ae} as the main base for image descriptor generation, the network can learn latent representations of the input image.

The main drawback of using the latent space of an \gls{ae} as a descriptor is that it considers every portion of the input image with the same importance. Furthermore, there is no control over the information represented in the latent space.

To solve this problem,adding an auxiliary classifier that shares the descriptor with \gls{ae} is proposed. This addition to the network will attempt to maintain the information from the patient's tumor. Here, it is important to note that this learning scheme keeps certain features of the input information, to disentangle the patient's healthy and tumoural information in the image descriptor. This solution is based on the work of~\cite{kobayashi2021decomposing} where three different \glspl{ae} were used to disentangle tumor and normal information from each case.

Using the classifier, the network is forced to learn the tumor's information to classify the cases where a tumor is present. At the same time, \gls{ae} forces the descriptor to maintain the structural characteristics of each patient. This dual-objective forces the network to focus on some present features in this case.

The main idea is to combine the ability of an \gls{ae} to represent in a small image descriptor the features of an image with the pathology detection capability of a classifier. The results presented in Section \ref{section:Results} show that the proposed architecture outperforms previous approaches in retrieving more similar images from a database for a given query. By combining the feature extraction potential of both \gls{ae} and classifier, the \gls{mocae} can improve each architecture individually and previous similar architectures.

This architecture is focused on medical \gls{cbir} because it makes it possible to make the network focus on the possible pathologies of each image. About the work of \cite{kobayashi2021decomposing}, this scheme does not require a segmented dataset. In the proposed method, the binary classifier only needs information about the presence or absence of a tumor in the image, but segmentation of the tumoural region of the image is not necessary.

Thus, the proposed model takes as an input a brain three dimensional \gls{mr} and extracts the most relevant patient information, regarding anatomical and pathological features. I.e. representing in a small vector the information of the healthy portion of the brain and, in case of presence, the tumor features. Then, this vector is used to compare each case with the documented cases of the database. The most similar cases are recommended to the physician, this way presenting the professional the most similar cases to support its diagnostic.

As stated in \cite{krishnan2022self}, in the field of medicine, the process of annotating the data is particularly costly because specialists in the field must perform manual annotation. Therefore, the possibility of developing a \gls{cbir} system capable of focusing on the most relevant areas of the image while maintaining the relatively low cost of the labels used is a characteristic that is particularly desired.

The project's source code, along with trained models and results are publicly available\footnote{\url{https://purl.com/mocae_brats}}.

\subsubsection{Model training scheme}
\label{section:ModelTraining}
The proposed architecture must combine two different learning processes using the same parameters. As stated above, this shared information forces the network to combine each image's normal and pathological features.

Regarding the model's training, two different outputs will use two different losses, which must be combined to train the model. The \gls{ae} output is responsible for reconstructing the information in the input image, while the classifier must differentiate between healthy and regions with the presence of a tumor in it.

First, the input image $x$ must be reconstructed by the \gls{ae} generating $\hat{x}$, both the input and the reconstructed image will be compared using the L2 norm.
\begin{equation}
    L_{r}(x, \hat{x}) = \vert \vert x-\hat{x} \vert \vert_{2} ^2
\end{equation}

This reconstruction loss function $L_{r}$ will force the latent space to learn the spatial features of the input image.

At the same time, the classification head of the proposed model will be trained using the binary cross-entropy loss function, formulated as follows:
\begin{equation}
    L_{c}(x)=-\frac{1}{N} \sum_{i=1}^N y_i \cdot \log (\hat{y}_i)+(1-y_i) \cdot \log (1-\hat{y}_i)
\end{equation}

The head loss classification function $L_{c}$ focuses on differentiating healthy and tumoural images and detecting the presence of tumors in the input image. When seeking this objective, the model will be forced to maintain tumoural features in the latent space.

Both losses are combined to train the model using the following equation:
\begin{equation}
    \label{equation:TotalLoss}
    L_{t} = \gamma_{1} * L_{r} + \gamma_{2} * L_{c}
\end{equation}
where both $\gamma$ terms are normalization coefficients to balance both losses.

The loss function is minimized using the Adam~\cite{kingma2014adam} optimizer. The best values of the $\gamma$ parameters were found using the grid search in the training phase. The range of values tested varied between $0$ and $1$, requiring that the sum of both $\gamma$ must be of $1$. The best performance was achieved with $\gamma_{1} = 0.2$ and $\gamma_{2} = 0.8$.

\subsubsection{Network architecture definition}
\label{section:NetworkArchitecture}
The proposed network composition is \gls{cnn} using the Residual blocks presented in the ResNet architecture~\cite{he2016deep}. The same principles as those used in ResNet were used to design the Encoder and Decoder networks of the model. In particular, it was decided to use complete pre-activation blocks as was proposed in~\cite{he2016identity}, that generally produce the best results. The \gls{ae} generates the latent vector by reducing the information to a space of 500 dimensions that corresponds to the image descriptor.

Regarding the classifier, it uses the latent vector information of 500 positions and generates a binary classification, using an intermediate dense layer of 64 neurons.

\begin{figure*}[h]
	\centering
	\includegraphics[width=\linewidth]{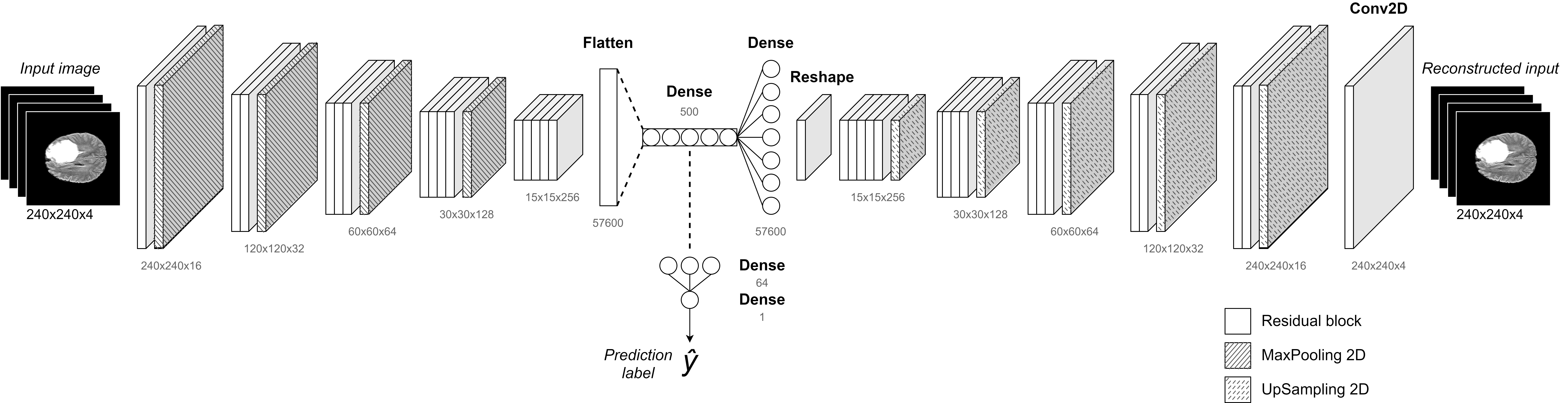}
    \caption{\gls{mocae} neural network detailed architecture.}
    \label{figure:ModelArchitecture}
\end{figure*}

Figure \ref{figure:ModelArchitecture} shows details of the model architecture. The input data that the Encoder receives are 4 slices of each case, corresponding to the different \glspl{mr} (\gls{t1}, \gls{t1gd}, \gls{t2}, \gls{flair}). Each residual block contains two separable convolutions~\cite{chollet2017xception} along with Batch Normalization~\cite{ioffe2015batch} and Dropout~\cite{hinton2012improving} layers. The \gls{relu} activation function was used in the hidden layers, while the hyperbolic tangent and the sigmoid activation were used in the reconstruction and classification outputs, respectively.

\subsubsection{\gls{cbir} algorithm}
\label{section:CBIRALgorithm}
\gls{cbir} system aims to obtain the patient's most similar to the query from the database of documented clinical cases. In other words, each time a \gls{mr} is received, the algorithm \gls{cbir} finds in the database the most similar cases in the database by comparing the healthy and tumoural structures of the query.

The image descriptor is generated to compare each query with the rest of the documented cases. This descriptor corresponds to the latent space of \gls{mocae} generated using the trained Encoder network. We use the Euclidean distance to compare the query image with the rest of the database, following the same approximation as~\cite{kobayashi2021decomposing, tarjoman2013implementation, shakarami2020efficient, siradjuddin2019feature}.

\begin{figure}
    \centering
    \includegraphics[width=0.8\textwidth]{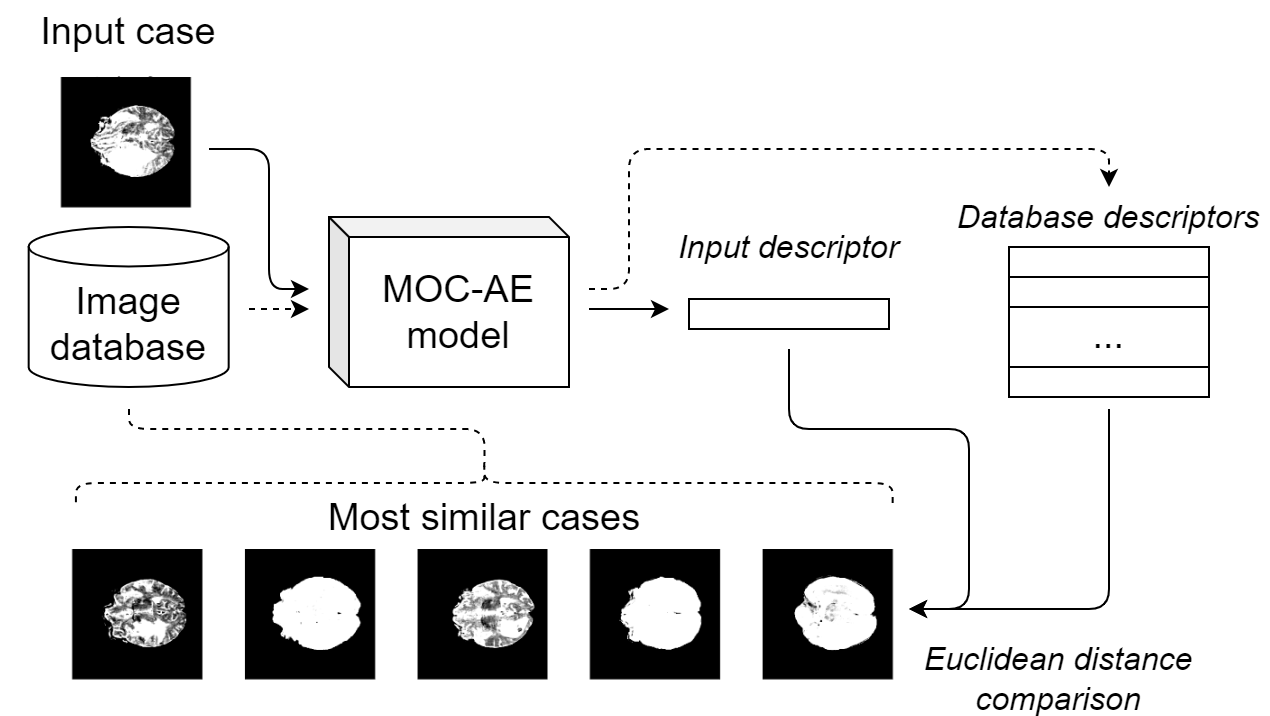}
    \caption{\gls{cbir} diagram of the proposed recommendation system.}
    \label{fig:CBIR_Diagram}
\end{figure}

Figure \ref{fig:CBIR_Diagram} contains a scheme of the recommendation system. The architecture takes an input patient and generates its corresponding feature descriptor with the proposed \gls{mocae} network. Then, the most similar cases are retrieved from the database by comparing the descriptors with the Euclidean distance.

The recommendation system will also use the classification learning of the network to retrieve similar cases from the database. This information will be used to enhance the recommendation because of the network’s inherent ability to classify a certain query as tumoural or not. If the query is classified as tumoural by the classifier with certain reliability it will only search among other tumoural images. This way, when the network is sure that a case contains a tumor it will only search for other tumoural cases. In particular, it is decided to put a threshold of 90\% of confidence from where search on tumoural cases.

\section{Results}
\label{section:Results}
The dataset from the input of the experiment consists of 369 cases divided into 155 slices each. Thus, a total of 57.195 images were used. However, 10\% of the whole dataset (5.720 images) is randomly reserved for testing purposes. This partition is made by randomly reserving 10\% of the patients for evaluation purposes

The results of proposed algorithm performance by its training results and by comparing it with similar works are described as follows. 

\subsection{Training results}
The network was trained on a NVIDIA Quadro RTX 8000. The network was trained during 25 epochs, achieving the best results on the epoch 20. The model weights using during the rest of the results corresponds with the best performance of the model, regarding the total loss described in equation \ref{equation:TotalLoss}.

The network training time for epoch was 1681$\pm$12 seconds with a maximum of 1681 seconds and a minimum of 1612 seconds. The batch size using during the experiment was 32, with a total of 1429 iterations each epoch.

Respect inference time, the model predicts both outputs in 0.0658$\pm$0.0406 seconds with a maximum inference time of 1.0154 seconds and a minimum of 0.0478 seconds.

We evaluate the model performance against the binary classification of tumor presence in each \gls{mr}. The classification output of the model was measured for the validation split of the dataset. That is, the model was tested using images never seen before. The results obtained, with respect to the confusion matrix, can be analyzed in Table \ref{table:ConfusionMatrix}.

\begin{table}[h]
\centering
\begin{tabular}{|cc|cc|}
\hline
\multicolumn{2}{|c|}{\multirow{2}{*}{Total images = 100}} & \multicolumn{2}{c|}{Predicted value} \\ \cline{3-4} 
\multicolumn{2}{|c|}{} & \multicolumn{1}{c|}{0} & 1 \\ \hline
\multicolumn{1}{|c|}{\multirow{2}{*}{Real value}} & 0 & \multicolumn{1}{c|}{49} & 8 \\ \cline{2-4} 
\multicolumn{1}{|c|}{} & 1 & \multicolumn{1}{c|}{8} & 35 \\ \hline
\end{tabular}
\caption{Confusion matrix of the \gls{mocae} classification output.}
\label{table:ConfusionMatrix}
\end{table}

On the other hand Table \ref{table:TrainingMetrics} includes different metrics for the binary classification output of the model.

\begin{table}[h]
\centering
\begin{tabular}{c|c|c|c|c|}
\cline{2-5}
                                   & precision & recall & f1-score & support \\ \hline
\multicolumn{1}{|c|}{0}            & 0.86      & 0.86   & 0.86     & 57      \\ \hline
\multicolumn{1}{|c|}{1}            & 0.81      & 0.81   & 0.81     & 43      \\ \hline
\multicolumn{1}{|c|}{accuracy}     &    -      &   -    & 0.84     & 100     \\ \hline
\multicolumn{1}{|c|}{macro avg}    & 0.84      & 0.84   & 0.84     & 100     \\ \hline
\multicolumn{1}{|c|}{weighted avg} & 0.84      & 0.84   & 0.84     & 100     \\ \hline
\end{tabular}
\caption{Classification metrics of the \gls{mocae}.}
\label{table:TrainingMetrics}
\end{table}

Respect the reconstruction output of the network, the results are measured using the mean squared error, the root mean square error and the mean absolute error, pixel by pixel in all cases, obtaining a mean square error of 0.02805, a root mean square error of 0.16749 and a mean absolute error of 0.05312. These values correspond with the reconstruction of the input patient \glspl{mr}.

\subsection{Recommendation results}
After analyzing the model's training performance, we test our model against comparative diagnostic tasks. The \gls{cbir} model's performance is evaluated first empirically and then quantitatively. These results show the comparative diagnostic model performance, once trained, when the most similar cases are retrieved for a particular patient.

\subsubsection{Empirical evaluation of the \gls{cbir} system}
We show the behavior of the model to empirically evaluate recommendation results. These experiments show how the model performs against different cases, we analyze it's results respect the medical diagnosis perspective, trying to figure out similarities and mismatches between the input and retrieved images.

\begin{figure}[h]
    \centering
	\includegraphics[width=\linewidth]{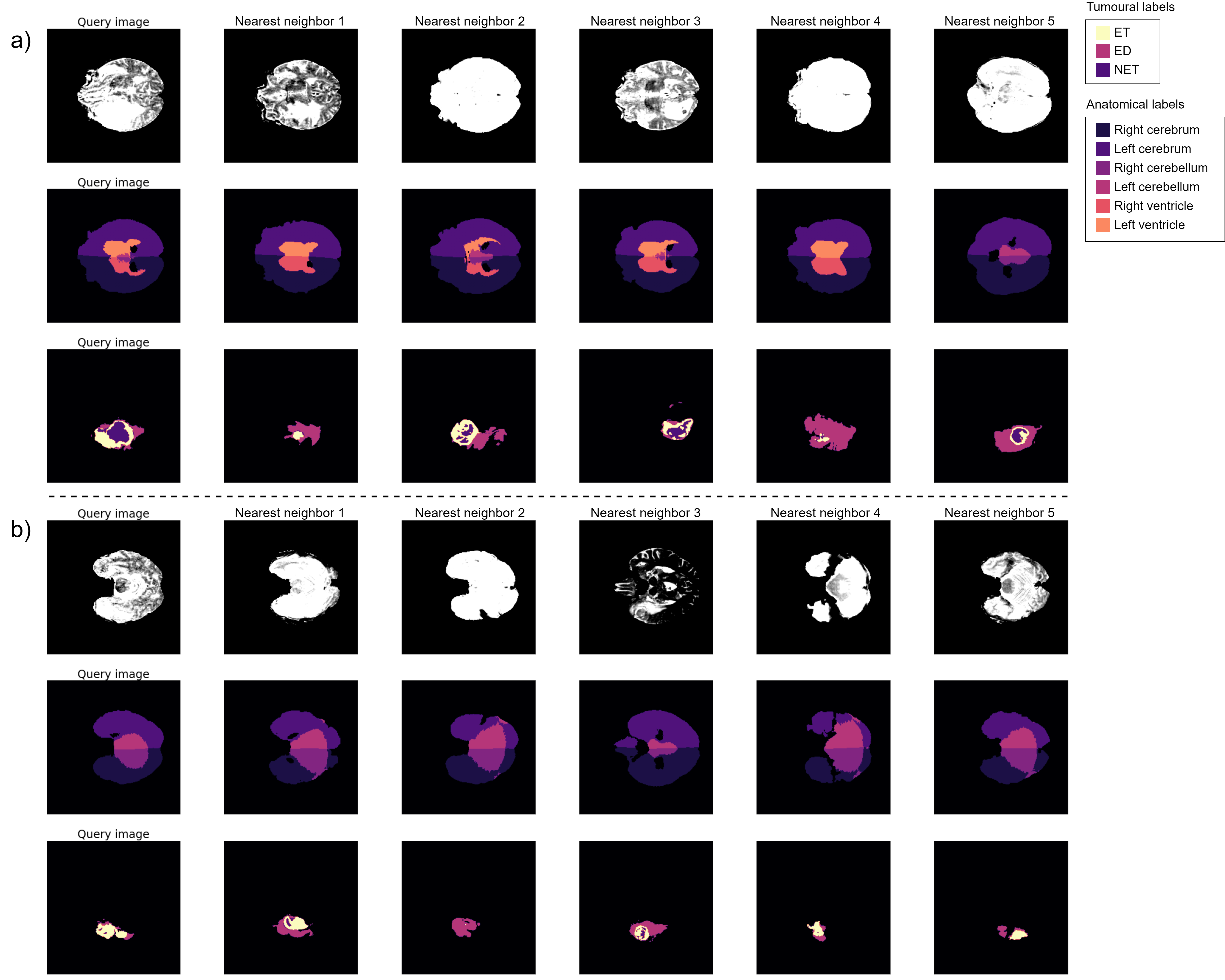}
    \caption{Nearest neighbors recovered by the \gls{mocae} for two different queries, ordered from left to right.}
    \label{figure:CBIRResults}
\end{figure}

\begin{figure}[h]
	\centering
	\includegraphics[width=\linewidth]{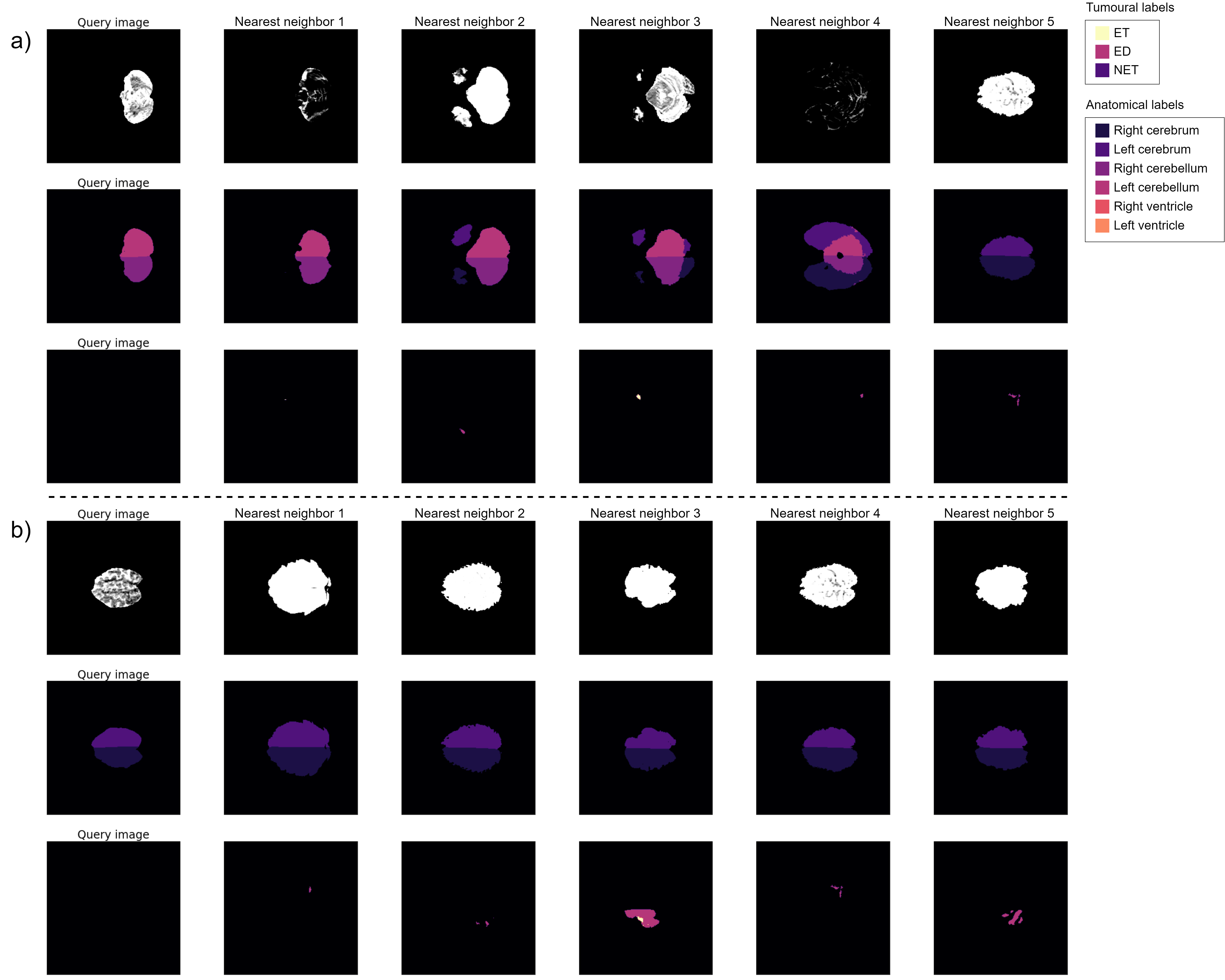}
    \caption{Nearest neighbors recovered by the \gls{mocae} for two different queries with absence of tumor, ordered from left to right.}
    \label{figure:CBIRResults1}
\end{figure}

To compare model's results Figure \ref{figure:CBIRResults} shows the results for two different queries used as input to the system. The first column of the image corresponds with the input query image that is being used for image retrieval. Then, the 5 more similar cases detected by the proposed system are showed, ordered by closeness from left to right. As can be seen, the model can retrieve similar images in both anatomical and tumoural aspects. At the same time, Figure \ref{figure:CBIRResults1} contains two cases of brain regions with no tumor present in it, also with the 5 more similar cases ordered by closeness from left to right. This second case of study shows the model's performance against normal cases.

Figure \ref{figure:CBIRResults}a represents a patient with a tumor in the middle section of the brain. The tumor is positioned in the right hemisphere of the cerebrum; the retrieved images from the dataset represent cases of the same brain region also sharing a tumor of similar composition with similar position and geometry.

On the other hand, Figure \ref{figure:CBIRResults}b shows a case with a smaller tumor in a lower region of the brain, positioned in the right cerebellum. The similarity between the query and the retrieved cases exhibits the performance of the model, being able to extract at the same time the anatomy of the patient, i.e. the region of the images is the same, at the same time that the tumor is well extracted and represented, i.e. the similarity and location between the tumor cases is very high.

Figure \ref{figure:CBIRResults1} follows the same structure as Figure \ref{figure:CBIRResults} but with non tumoural sections. Figure \ref{figure:CBIRResults1}a shows a case of a lower section of a patient's brain with no tumor present in it. As seen, the model retrieves from the database cases from the same section anatomically, i.e. cases share the same brain region and shape. Respect the tumor correlation between images, the results represent cases without a tumor present or tumors with small area.

In the case represented in Figure \ref{figure:CBIRResults1}b it can be seen as a case of the brain's upper area without a tumor. Similar to the case \ref{figure:CBIRResults1}a the retrieved images share the absence or presence of very small tumor while presenting a case of a similar brain area.

The results show that when there is no tumor present in the query, the retrieved cases have small or non-tumors. This highlights the need for a more diverse dataset for training, as our current dataset exclusively contains cases with tumors. While this composition provides valuable information on tumor cases, it limits the model's ability to accurately identify non-tumor cases due to the low number of such cases. To improve upon this, future work could focus on diversifying the dataset to include a balanced representation of both tumor and non-tumor cases, or apply data augmentation techniques to simulate a more diverse dataset. Cross-validation using different datasets could also be considered to add robustness to our findings.

\subsubsection{Similarity results of the recommendation}
To further evaluate the \gls{cbir} potential of our model we decide to quantitatively analyze it against other similar research. We evaluate the performance of the comparative model against Kobayashi et al research~\cite{kobayashi2021decomposing}. These results evaluate how similar are the retrieved images of the database with respect the input image being diagnosed, therefore providing a measure of the similarity of the retrieval.

Table~\ref{table:KobayashiComparison} compares of the performance of the proposed model with that of the model developed by Kobayashi et al~\cite{kobayashi2021decomposing} in terms of the Sørensen-Dice coefficient~\cite{dice1945measures, sorensen1948method}. In particular, the table measures the consistency between the anatomy of the healthy and tumoural features of the image. This coefficient varies between 1 and 0, where the higher the value, the more similar are the query and the retrieved images. To calculate the Sørensen-Dice coefficient of the proposed method, the same procedure will be followed as~\cite{kobayashi2021decomposing}, where the queries used are the slices of each case with the largest tumoural area. Three different coefficients are analyzed, (i) the \textbf{Normal Dice}, which evaluates the model's ability to identify and retrieve images based on their normal or healthy characteristics, using the six anatomical labels obtained with the procedure explained in Section \ref{section:DataPreprocessing} and calculating the multi-label Sørensen-Dice coefficient, (ii) the \textbf{tumoural Dice} which assesses the model's skill in identifying and retrieving images based on their tumoural or pathological features uses the segmentation information of each image, measuring the similarity of the tumoural sections of each case, and (iii) the \textbf{Entire Dice} which measures the model's overall proficiency in retrieving relevant images considering both normal and tumoural characteristics, thus providing a more complete measurement. We set the number of neighbors of the \gls{cbir} to 5, following the same procedure as~\cite{kobayashi2021decomposing}. The Entire Dice corresponds with the mean value between the normal and tumoural Dices.

\begin{table*}[h]
\centering
\begin{tabular}{|l|c|c|c|}
\hline
\multicolumn{1}{|c|}{Model} & Normal Dice & Tumoural Dice & Entire Dice \\ \hline
\gls{mocae} & 0.632$\pm$0.218 & \textbf{0.316}$\pm$\textbf{0.275} & \textbf{0.474}$\pm$\textbf{0.173} \\ \hline
\begin{tabular}[c]{@{}l@{}}Kobayashi et al.\\  Normal latent space\end{tabular} & \textbf{0.730}$\pm$\textbf{0.196} & 0.072$\pm$0.098 & 0.401$\pm$0.108 \\ \hline
\begin{tabular}[c]{@{}l@{}}Kobayashi et al.\\  Abnormal latent space\end{tabular} & 0.505$\pm$0.235 & 0.289$\pm$0.120 & 0.397$\pm$0.137 \\ \hline
\begin{tabular}[c]{@{}l@{}}Kobayashi et al.\\  Entire latent space\end{tabular} & 0.695$\pm$0.208 & 0.201$\pm$0.158 & 0.448$\pm$0.123 \\ \hline
\end{tabular}
\vspace*{5mm}
\caption{Sørensen-Dice coefficient values comparing \gls{mocae} and the work of \cite{kobayashi2021decomposing}.}
\label{table:KobayashiComparison}
\end{table*}

The performance of our model in identifying and retrieving images based on their normal or healthy characteristics yields a Normal Dice score of $0.632 \pm 0.218$. Furthermore, \gls{mocae}'s Tumoural Dice score stands at 0.316±0.275. The overall proficiency of \gls{mocae}, encapsulated by the Entire Dice score and considering both normal and tumoural characteristics, registers at 0.474±0.173.

Compared with Kobayashi et al.'s model, it's evident that while their work may slightly surpass \gls{mocae} in retrieving normal features, our contribution significantly outperforms in the identification and retrieval of tumoural features from medical images. This capability is particularly valuable, especially considering the primary goal of our study and the role of the \gls{cdss} in aiding the diagnosis of tumoural pathologies.

Moreover, a critical strength of the presented model lies in its ability to train the network without needing segmentation information for the cases, setting it apart from the model of Kobayashi et al. Although the best anatomical similarity, measured in the Normal Dice, is achieved by Kobayashi et al., the overall performance in terms of patient tumoural features shows noticeable improvement with \gls{mocae}.

Consequently, the proposed model efficiently manages the delicate equilibrium between each patient's healthy and tumoural areas, enabling the retrieval of highly similar images from the database. It presents an optimal balance between normal and pathological features in patients. It also demonstrates superior performance according to state-of-the-art standards while concurrently reducing the cost and complexity inherent in the training process. These characteristics highlight \gls{mocae}'s promising potential for use in the medical field.

\section{Discussion}
\label{section:Discussion}
\glspl{cdss} represent critical algorithms that could significantly improve the diagnostic tasks of doctors. The proposed model achieves state-of-the-art results in both the healthy and tumoural features of recommended cases. The model can factorize images focusing on both the normal features of the patient and the pathologies present in the case, generating a compact and significant representation of each image. These image descriptors can then be used to recommend similar cases from a database, helping the physician to make the diagnosis.

The architecture presented in the current work can combine features present in the image with labels annotated by professionals. One of the main strengths differentiating \gls{mocae} from similar \gls{cbir} models is that it does not require costly information on the label such as tumor segmentation. Our model is trained only using the binary labels of presence or absence of a tumor, while the costly segmentation labels are only used during evaluation of the model's performance. The model learns characteristics of each patient combining recommendation and classification outputs, thus generating an enriched image descriptor using only binary label information. In areas where the cost of producing high-quality labels is especially costly, it is considered crucial to develop a model that can produce state-of-the-art results with a low-cost associated.

In contrast to previous works~\cite{bhalodiya2022magnetic, chen2022efficient, ghaffari2022automated} our model does not require segmentation labels. Tumor area segmentation is a costly information and in many cases is very difficult to obtain, it is considered that it is far more common to be available information of the presence or absence of a tumor in an image. Thus, the proposed architecture is considered an alternative that can be extended to a wider range of cases, especially in the medical area where data availability is difficult. The only necessary information to apply our system to another case of study are the images that will be ranked along with their corresponding binary labels. In our case, the segmentation labels are used only for evaluation purposes and could be adapted to evaluate the new target domain. This is of great interest, not only for facilitating its use and implementation but also because it can be generalized to every medical area due to a cost reduction. The simple general structure that we provide opens up the possibility to apply the proposed system to new domains, with almost any change.

Previous researches such as \cite{lehmann2005automatic}  recommend images using the nearest neighbor algorithm, as was done in other research independently of medical imaging~\cite{siradjuddin2019feature, shakarami2020efficient, lehmann2005automatic}. Respect this work, our model is based on \gls{cnn}, that have emerged as the state-of-the-art in \gls{cv} tasks \cite{bhatt2021cnn}.

The work of \cite{tarjoman2013implementation} proposes a simple feature extraction method using \gls{svm}~\cite{cortes1995support} and the Grey Level Co-occurrence Matrix as the main input of each image. A similar approximation was followed in \cite{kumar2016adapting}, where \gls{svm} is used as the descriptor generator mechanism, in this case using different features of the liver images. Finally, using the weighted nearest neighbor~\cite{cover1967nearest} a query classification is produced. The presented architecture differs from this approximation in the usage of the whole image as the input for the model. By using the Grey Level Co-occurrence Matrix in \cite{tarjoman2013implementation} the textural information of the image is maintained, but it could lead to huge information loss. Using the complete image provides our model all the information from where it can learn the relevant features of each image.

The work presented in~\cite{kobayashi2021decomposing} proposes a \gls{cbir} scheme based on \glspl{ae} to extract the most important features of brain tumor images. This work uses three different \gls{ae} that generate three different image descriptors, one focused on the healthy features of the image, the other on the tumor area, and the last one uses the information from the entire image. Using these different outputs, researchers can disentangle the normal and abnormal characteristics of the query to provide a controlled recommendation. This work combines \gls{dl} techniques and traditional medical \gls{cbir} to recommend similar database images given a certain query. The architecture proposed in this research is dependent on the availability of segmentation labels, which can be difficult to obtain. Our work differs from this by using less costly labels opening up the possibility of applying the model to new cases.

The results show that the presented architecture can extract from the brain \glspl{mr} information about the anatomy of the patient and the presence and composition of tumors simultaneously, outperforming previous solutions. Respect \cite{kobayashi2021decomposing}, there is an improvement in Sørensen-Dice coefficient from 0.448 to 0.474 while reducing the cost of the training labels. The recommendations suggest that the inclusion of the classification output enriches the image descriptor, while not losing structural information of the anatomy of the patient. It is specially interesting to observe that, although anatomical information of the patients is not available, the model can recommend cases with similar healthy structures.

Summarizing, the model's results improve the previous work by obtaining better results in both retrieving cases with similar pathologies and balancing both anatomical and abnormal features of each case. Given the results, \gls{mocae} is considered the best alternative to develop a \gls{cdss} specialized in medical imaging recommendation. In this sense, this AI model could become a helpful tool for clinicians to make more accurate image diagnoses and make more proper decisions avoiding some subjective biases.

Overall, the main contributions of this paper are summarized as follows:
\begin{itemize}
    \item The \gls{mocae} architecture, a \gls{dl} model for image descriptor generation, is presented. The network is specifically designed for medical image recommendation, but its simplicity makes it possible to extend it a broader variety of cases.

    \item The design of the network is discussed, focusing on alleviating the cost of the data used to train the network while improving the performance of the recommendation.

    \item Experiments showcasing and comparing the model are presented. The results show that our contribution outperform previous networks in retrieving more similar cases respect tumoural features and overall similarity of the brain \glspl{mr}.
\end{itemize}

The presented model has several aspects that could be improved in further researches. First, the model's performance evaluation is only defined by using the segmentation labels. It is not correct to only evaluate the correlation between labels, because one of the strengths of the model is that it can link very similar cases that may or may not share the same pathological features and be very similar visually. This characteristic is crucial because it provides to the physician additional information for a given case, making possible to identify new pathologies with this similarity. To only evaluate the similarity if the images (e.g. by comparing the value of their pixels) is also not correct, because this measurement will not take into account the similarity of the cases in medical terms. Therefore, further research should focus on defining new methods of evaluation, considering all the important aspects of CBIR.

Second, the model performance could be improved by applying optimization network methods, e.g. further tunning the hyperparameters of the neural network or adding attention layers to the network. 

Thirdly, besides lowering the cost of obtaining the dataset, the architecture still need labels for train. This requirement could limit the range of applications of the framework, in particular, in problems where there is not labeled data it is not possible to use our model. Besides that, it should be noted that the segmentation information used in our research is only used for evaluation purposes, and in new domains can be omitted with different evaluation metrics. Further research must be carried out to evaluate the architecture's performance against different cases of study.

%\section{Conclusions}
%\label{section:Conclusions}
Using the proposed model in tumoural \gls{cbir} arises a promising option to improve efficiency and accuracy in comparative diagnostic and tumoural pathologies treatment, with the possibility of helping physicians make more accurate diagnosis. Furthermore, the proposed framework has the flexibility to be applied to different medical diagnostic cases.

\bibliographystyle{elsarticle-num}
\bibliography{refs.bib}

\section*{Declaration of Competing Interest}
The authors declare that they have no known competing financial interests or personal relationships that could have appeared to influence the work reported in this paper.

\section*{Acknowledgements}
The authors thank the research group KNOwledge Discovery and Information Systems (KNODIS) for the compute infrastructure.

\section*{Supplementary Material}
The source code of the project, along with trained models and results are publicly available and can be consulted in \url{https://purl.com/mocae_brats}.

\end{document}